\documentclass[twocolumn,english,prl]{revtex4}
\usepackage[latin9]{inputenc}
\usepackage{textcomp}
\usepackage{bm}
\usepackage{amsmath}
\usepackage{graphicx}
\usepackage{amssymb}
\usepackage{esint}

\makeatletter
\@ifundefined{textcolor}{}
{%
 \definecolor{BLACK}{gray}{0}
 \definecolor{WHITE}{gray}{1}
 \definecolor{RED}{rgb}{1,0,0}
 \definecolor{GREEN}{rgb}{0,1,0}
 \definecolor{BLUE}{rgb}{0,0,1}
 \definecolor{CYAN}{cmyk}{1,0,0,0}
 \definecolor{MAGENTA}{cmyk}{0,1,0,0}
 \definecolor{YELLOW}{cmyk}{0,0,1,0}
 }

\makeatother

\makeatother

\usepackage{babel}

\begin{document}

\title{Quantum System Identification}

\author{Daniel Burgarth$^{1}$ and Kazuya Yuasa$^{2}$}

\address{$\text{\textonesuperior}$QOLS, Blackett Laboratory, Imperial College London, London, SW7 2BW, United Kingdom}

\address{$\text{\texttwosuperior}$Waseda Institute for Advanced Study, Waseda University, Tokyo 169-8050, Japan}
\begin{abstract}
%The aim of quantum system identification is to estimate the Hamiltonian, the state, the control fields, and the measurement observables of a quantum system. Here we establish a criterion that allows us to classify how much knowledge about the system can be extracted, in principle, from a given experimental setup. We show how the structure of observables plays an important role in the estimateion efficiency and provide a method to estimate arbitrary Hamiltonians provided the system topology is known.
The aim of quantum system identification is to estimate the ingredients inside a black box, in which some quantum-mechanical unitary process takes place, by just looking at its input-output behavior.
Here we establish a basic and general framework for quantum system identification, that allows us to classify how much knowledge about the quantum system is attainable, in principle, from a given experimental setup.
Prior knowledge on some elements of the black box helps the system identification.
We present an example in which a Bell measurement is more efficient to identify the system.
When the topology of the system is known, the framework enables us to establish a general criterion for the estimability of the coupling constants in its Hamiltonian.
\end{abstract}
\maketitle
Some of the most exciting and puzzling concepts in quantum theory
can already be observed in simple systems. These are, for example,
superpositions and decoherence, tunneling, entanglement and non-locality,
quantum cryptography, teleportation and dense coding. Many of such
theoretical ideas have been confirmed experimentally with a tremendous
accuracy. On the other hand, perhaps the most important theoretical
concept---a full quantum computer or simulator---is still well out
of reach, because it requires a fully controllable system of Hilbert
space dimension at the very least of the order of $2^{100}$. Its
realization poses one of the greatest challenges in science today.

On our path towards quantum computation we are building systems composed
of more and more qubits, the quantum information theoretic equivalent
of the bit. But while an information theoretic approach is very successful,
we should not forget that any implementation comes with a baggage
of physical effects. In particular, real qubits \emph{interact}. Often,
these interactions are important: they are actively used to create
logical gates. Sometimes, they are unwanted, and either suppressed
actively, or simply neglected. However if we are to meet the stringent
bounds that fault-tolerance computation puts on the required precision
of our technology, we will have to estimate our quantum system with
very high precision. Current estimates of the fault-tolerance threshold
indicate that in many systems the relative precision will have to
be of the order of $10^{-3}$--$10^{-4}$.

If we could perfectly control our system, achieving such precisions
is a mere engineering difficulty. But if our control relies on the
system couplings, or is heavily perturbed by them, we are in a Catch-22
situation, and it is unclear how well the system can be estimated
even in principle. In this paper, we solve this question by providing
a precise mathematical description of the \emph{equivalent set} \cite{ref:Sontag,ref:FAlbertiniDAlessandro,Albertini2008} of
closed systems. This set describes the possible implementations of
a system that cannot be distinguished with a given experimental setup.
It should be compared to the well-known reachable set in quantum control \cite{D'Alessandro2008},
which describes the set of unitary operations that can be implemented,
in principle, by a given experimental setup.

It has been shown in quantum control that even when only parts of
the system are accessed, the reachable set typically remains
maximal: the system is capable of quantum computation \cite{ref:Lloyd}.
We show that this is not true for full estimability: in general, infinitely
many different system Hamiltonians give rise to the same input-output
behavior. However, we show how a priori knowledge about the system
helps to restrict the set of possible systems. Indeed we prove that
in a generic limited-access situation, relatively little a priori
knowledge can imply full estimability. This generalizes several recently
developed schemes for indirect estimation \cite{Burgarth2009,Burgarth2009a,Franco2009,Burgarth2011}.
We also show how estimability can strongly depend on the structure
of quantum measurements, by providing an example where entangled observables
are more efficient for the estimation than product observables.

Our analysis first follows closely the known results from bilinear
theory \cite{ref:Sontag}. Then, we use a result from Lie algebras \cite{ref:LieAlgebras-Jacobson}
to translate the bilinear theory to the quantum case. This sets our
result apart from previous work which required additional mathematical
assumptions \cite{ref:FAlbertiniDAlessandro,Albertini2008}.
\begin{figure}[b]
\includegraphics[width=0.9\columnwidth]{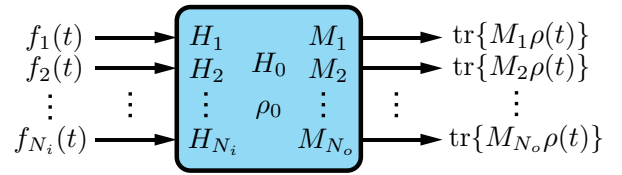}
\caption{\label{fig:Black-box-setup.}
%Black box setup. The inputs are functions that determine the time dependence of the system, while the outputs are expectation values of observables on the solution of the Schr\"odinger equation. 
A set of time dependent functions $f_k(t)$ is the input, which determines the unitary dynamics inside the black box, and a set of the expectation values of observables $M_\ell$ is the output.
Our objective is to estimate the system $\sigma=\{H_0,H_k,M_\ell,\rho_0\}$ by looking at the input-output behavior of the black box.
In the most extreme case, even the control operations $H_{k}$ and the observables $M_{\ell}$ are unknown.}
\end{figure}

\paragraph*{Setup.---}

We consider a black box with $N_{i}$ inputs and $N_{o}$ outputs.
Inside the black box, some quantum-mechanical unitary dynamics takes
place. Our goal is to find a model for the black box that perfectly
describes its input-output behavior under all possible circumstances
(\emph{system identification} \cite{ref:Sontag}).

More specifically, we are modeling a system with a finite-dimensional
Hilbert space $\mathcal{H}$, a time dependent Hamiltonian
\[
H(t)=H_{0}+\sum_{k=1}^{N_{i}}f_{k}(t)H_{k},
\]
an initial quantum state $\rho_{0}$, and a set of observables $M_{\ell}$
($\ell=1,\ldots,N_{o}$). Without loss of generality we chose $H_{0}$
and $H_{k}$ traceless. The inputs are the functions $f_{k}(t)$ ($k=1,\ldots,N_{i}$),
which are assumed to be piecewise constant. The outputs are the expectation values
of the observables $M_{\ell}$, 
\[
\mbox{tr}\{M_{\ell}\rho(t)\}\quad\text{with}\quad\rho(t)=T_{\rightarrow}\exp\!\left(\int_{0}^{t}dt'\,\mathcal{L}(t')\right)\rho_{0},
\]
where \[
\mathcal{L}(t)=\mathcal{L}_{0}+\sum_{k=1}^{N_{i}}f_{k}(t)\mathcal{L}_{k},\ \ \mathcal{L}_{k}=-i[H_{k},\bullet\,]\ \ (k=0,\ldots,N_{i})\]
 are the Liouvillians corresponding to the Hamiltonians.
See Fig.\ \ref{fig:Black-box-setup.}.
Because we are interested in whether systems can be distinguished
\emph{in principle}, we assume that it is possible to collect statistics
at arbitrary precision, and that infinitely many copies of the system
are available (this allows us to ignore any back-action of the measurements \cite{ref:FAlbertiniDAlessandro,Albertini2008}).
Our main assumption is that the system is controllable, implying that
any unitary transformation can be realized by the Hamiltonian dynamics
with $H(t)$, by properly arranging the inputs $f_{k}(t)$. Mathematically
this amounts to the smallest Lie algebra over the reals that contains
the matrices $iH_{0},iH_{1},\ldots,iH_{N_{i}}$ being equal
to the full Lie algebra $\mbox{su}(\dim\mathcal{H})$ of traceless
skew-hermitian matrices of size $\dim\mathcal{H}\times\dim\mathcal{H}$.
Controllability is a generic property of systems, and is in principle
an observable property, if the dimension of the underlying Hilbert
space is known. Furthermore we exclude the trivial cases where $M_{\ell}$
or $\rho_{0}$ is proportional to the identity operator.

We put all parameters together in the \emph{system} $\sigma=\{H_{0},H_{k},M_{\ell},\rho_{0}\}$.
Two systems $\sigma$ and $\hat{\sigma}$ are called \emph{equivalent}
\cite{ref:Sontag}, if they are indistinguishable by
all input-output experiments. Therefore by definition, we can estimate
the real system $\sigma$ up to equivalence. Let us call the estimated
system $\hat{\sigma}$, which consists of estimated components $\hat{\sigma}=\{\hat{H}_{0},\hat{H}_{k},\hat{M}_{\ell},\hat{\rho}_{0}\}$.
We assume that the estimated system has been chosen to be of minimal
dimension, which implies that this system is also controllable. The
goal is now to find a mathematical description of how different $\hat{\sigma}$
can be from the real system $\sigma$.

\paragraph*{Equivalence and similarity:---}

We first have to find a mathematical description of the equivalence.
For some fixed input, equivalence means that the real system and the
estimated system have to agree on all observable outputs for all times,
i.e., \begin{equation}
\mbox{tr}\{M_{\ell}\rho(t)\}=\mbox{tr}\{\hat{M}_{\ell}\hat{\rho}(t)\},\label{eqn:Equivalent}\end{equation}
 where $\hat{\rho}(t)$ is the state evolving from the initial state
$\hat{\rho}_{0}$ with the Hamiltonians $\hat{H}_{0}$ and $\hat{H}_{k}$.
This is not very useful mathematically, because it still involves
solving the Schr\"odinger equation. There is an algebraic description
of this property that is much easier. Let us denote $\mathcal{L}_{\bm{\alpha}}\equiv\mathcal{L}_{\alpha_{L}}\cdots\mathcal{L}_{\alpha_{1}}$,
where $\bm{\alpha}$ is a multi-index of length $L$ with entries
$\alpha_{j} \in {0,\ldots,N_{i}}$. Further, we include the case $L=0$
as the identity superoperator and introduce similar notation $\hat{\mathcal{L}}_{\bm{\alpha}}$ for the
estimated system. Equivalence
can then be formulated as \begin{equation}
\mbox{tr}\{M_{\ell}\mathcal{L}_{\bm{\alpha}}\rho_{0}\}=\mbox{tr}\{\hat{M}_{\ell}\hat{\mathcal{L}}_{\bm{\alpha}}\hat{\rho}_{0}\}\label{eqn:Equivalence}\end{equation}
 for any sequence of the indices $\bm{\alpha}$. This can be thought
of as an {}``infinitesimal version'' of (\ref{eqn:Equivalent}),
and a simple proof of this statement is found in \cite{ref:FAlbertiniDAlessandro}.

We call systems \emph{similar} if and only if there is a similarity
transformation between them\begin{equation}
\mathcal{L}_{k}=\mathcal{T}\hat{\mathcal{L}}_{k}\mathcal{T}^{-1},\quad\mathcal{M}_{\ell}=\hat{\mathcal{M}}_{\ell}\mathcal{T}^{-1},\quad\rho_{0}=\mathcal{T}\hat{\rho}_{0},
\label{eqn:Similarity}\end{equation}
$(k=0,\ldots,N_{i})$ where $\mathcal{M}_{\ell}$ and $\hat{\mathcal{M}}_{\ell}$ represent
the actions of $M_{\ell}$ and $\hat{M}_{\ell}$ in the Liouville
space. It is obvious that similarity implies equivalence. Similarity
is much easier to handle than equivalence, because of its simple mathematical
structure.

\paragraph*{Translation %of bilinear result 
to quantum case:---}

In bilinear system theory \cite{ref:Sontag} it was shown that
if $\sigma$ is controllable, then equivalence implies similarity.
This is proven by explicit construction of the similarity transformation
between $\sigma$ and $\hat{\sigma}$. Because there are some subtle
differences in the quantum case, we briefly repeat these arguments.

Assume $\sigma$ and $\hat{\sigma}$ are equivalent and pick an arbitrary
state $\hat{\rho}$. We show that due to the controllability of system
$\hat{\sigma}$, the state $\hat{\rho}$ can be expressed as \begin{equation}
\hat{\rho}=\sum_{\bm{\alpha}}\lambda_{\bm{\alpha}}\hat{\mathcal{L}}_{\bm{\alpha}}\hat{\rho}_{0}.\label{eqn:LiouvillianExpansion}\end{equation}
 Firstly, because $i\hat{\rho}_{0}\in\mbox{u}(\dim\mathcal{H})$ (the algebra of skew-hermitian matrices) the
set $R\equiv\{iA\,|\, A=\sum_{\bm{\alpha}}\lambda_{\bm{\alpha}}\hat{\mathcal{L}}_{\bm{\alpha}}\hat{\rho}_{0}\}$
is a subset of $\mbox{u}(\dim\mathcal{H})$. Because $i[\hat{\mathcal{L}}_{k},\hat{\mathcal{L}}_{j}]=-i\bm{[}[\hat{H}_{k},\hat{H}_{j}],\,\bullet\,\bm{]}$
and we have controllability, the linear combinations of $\hat{\mathcal{L}}_{\bm{\alpha}}$
include $\mathcal{L}_{\hat{H}}\equiv-i[\hat{H},\,\bullet\,]$ for
any hermitian $\hat{H}$. This means that $[i\hat{H},iA]\in R$, so
$R$ is an ideal. Because it is not equal to the identity and not $\mbox{su}(\dim\mathcal{H})$,
we must have $R=\mbox{u}(\dim\mathcal{H})$. Therefore, we can express
any hermitian operator as $\sum_{\bm{\alpha}}\lambda_{\bm{\alpha}}\hat{\mathcal{L}}_{\bm{\alpha}}\hat{\rho}_{0}$,
and in particular any state $\hat{\rho}$, as in (\ref{eqn:LiouvillianExpansion}).

We then define $\mathcal{T}$ by
\[
\mathcal{T}\hat{\rho}=\sum_{\bm{\alpha}}\lambda_{\bm{\alpha}}\mathcal{L}_{\bm{\alpha}}\rho_{0}.
\]
There are many possible representations of $\hat{\rho}$. In order
to see that $\mathcal{T}$ is well-defined as a mapping, we need to
verify that any two equal representations $\sum_{\bm{\alpha}}\lambda_{\bm{\alpha}}\hat{\mathcal{L}}_{\bm{\alpha}}\hat{\rho}_{0}=\sum_{\bm{\alpha}}\lambda_{\bm{\alpha}}'\hat{\mathcal{L}}_{\bm{\alpha}}\hat{\rho}_{0}$
imply $\sum_{\bm{\alpha}}\lambda_{\bm{\alpha}}\mathcal{L}_{\bm{\alpha}}\rho_{0}=\sum_{\bm{\alpha}}\lambda_{\bm{\alpha}}'\mathcal{L}_{\bm{\alpha}}\rho_{0}$.
By linearity, it is enough to show that \begin{equation}
\sum_{\bm{\alpha}}\lambda_{\bm{\alpha}}\hat{\mathcal{L}}_{\bm{\alpha}}\hat{\rho}_{0}=0\quad\Longrightarrow\quad\sum_{\bm{\alpha}}\lambda_{\bm{\alpha}}\mathcal{L}_{\bm{\alpha}}\rho_{0}=0.\label{eqn:Zero}\end{equation}
 Suppose that $\sum_{\bm{\alpha}}\lambda_{\bm{\alpha}}\hat{\mathcal{L}}_{\bm{\alpha}}\hat{\rho}_{0}=0$.
Then, we have for any $\bm{\beta}$ \[
\mbox{tr}\biggl\{\hat{M}_{\ell}\hat{\mathcal{L}}_{\bm{\beta}}\sum_{\bm{\alpha}}\lambda_{\bm{\alpha}}\hat{\mathcal{L}}_{\bm{\alpha}}\hat{\rho}_{0}\biggr\}=\sum_{\bm{\alpha}}\lambda_{\bm{\alpha}}\,\mbox{tr}\{\hat{M}_{\ell}\hat{\mathcal{L}}_{\bm{\beta}\bm{\alpha}}\hat{\rho}_{0}\}=0,\]
 where $\hat{\mathcal{L}}_{\bm{\beta}\bm{\alpha}}=\hat{\mathcal{L}}_{\bm{\beta}}\hat{\mathcal{L}}_{\bm{\alpha}}$.
Now, due to the input-output equivalence (\ref{eqn:Equivalence})
between the two systems $\sigma$ and $\hat{\sigma}$, i.e., $\mbox{tr}\{\hat{M}_{\ell}\hat{\mathcal{L}}_{\bm{\beta}\bm{\alpha}}\hat{\rho}_{0}\}=\mbox{tr}\{M_{\ell}\mathcal{L}_{\bm{\beta}\bm{\alpha}}\rho_{0}\}$
for any $\bm{\beta}\bm{\alpha}$, we get \[
\mbox{tr}\biggl\{ M_{\ell}\mathcal{L}_{\bm{\beta}}\sum_{\bm{\alpha}}\lambda_{\bm{\alpha}}\mathcal{L}_{\bm{\alpha}}\rho_{0}\biggr\}=\sum_{\bm{\alpha}}\lambda_{\bm{\alpha}}\,\mbox{tr}\{M_{\ell}\mathcal{L}_{\bm{\beta}\bm{\alpha}}\rho_{0}\}=0.\]
 Since this holds for any $\bm{\beta}$ and the system is controllable,
we conclude $\sum_{\bm{\alpha}}\lambda_{\bm{\alpha}}\mathcal{L}_{\bm{\alpha}}\rho_{0}=0$,
which completes the proof of (\ref{eqn:Zero}). The mapping is onto
due to the controllability of the system, and is shown to be one-to-one
by reversing the argument which proved that it is well defined. Finally
using controllability and the property $\mathcal{T}\hat{\rho}_{0}=\rho_{0}$
it is easy to see that $\mathcal{T}$ has to fulfill (\ref{eqn:Similarity}).

\paragraph*{%Proof of 
Unitarity:---}

Since we restrict ourselves to unitary dynamics, it is possible
to prove that the above similarity $\mathcal{S}(\,\bullet\,)\equiv\mathcal{T}\bullet\mathcal{T}^{-1}$
is actually inducing a unitary transformation on Hamiltonians. First,
we note that both the real and the estimated Liouvillians have the
commutator structure $\mathcal{L}_{k}=-i[H_{k},\bullet\,]$ and $\hat{\mathcal{L}}_{k}=-i[\hat{H}_{k},\bullet\,]$,
because we restrict ourselves to unitary dynamics. These Liouvillians
form a subspace $\mathcal{U}$ of all possible Liouvillians. We first
show that controllability implies that this subspace is mapped into
itself by the similarity transformation $\mathcal{S}$. Indeed, a
simple expansion of commutators combined with $\mathcal{S}$ being
a similarity transformation shows that \begin{align}
\mathcal{S}(-i\bm{[}[\hat{H}_{k},\hat{H}_{j}],\bullet\,\bm{]})& =\mathcal{S}(i[\hat{\mathcal{L}}_{k},\hat{\mathcal{L}}_{j}])=i[\mathcal{L}_{k},\mathcal{L}_{j}]\nonumber \\
 & \qquad\ \ =-i\bm{[}[H_{k},H_{j}],\bullet\,\bm{]} \in\mathcal{U}.\label{eq:liemorph}\end{align}
%-i\bm{[}[H_{k},H_{j}],\bullet\,\bm{]} & =i[\mathcal{L}_{k},\mathcal{L}_{j}]=\mathcal{S}(i[\hat{\mathcal{L}}_{k},\hat{\mathcal{L}}_{j}])\nonumber \\
% & \qquad\ \ =\mathcal{S}(-i\bm{[}[\hat{H}_{k},\hat{H}_{j}],\bullet\,\bm{]})\in\mathcal{U}.\label{eq:liemorph}\end{align}
 By linearity, any element $\hat{H}$ of the generated algebra has
the property that $\mathcal{S}(-i[\hat{H},\bullet\,])\in\mathcal{U}$.
Because the system is controllable, this algebra is just the set of
all traceless hermitian matrices, and therefore $\mathcal{S}(\mathcal{U})=\mathcal{U}$.
Since there is an isomorphism between $\mathcal{L}_{\hat{H}}=-i[\hat{H},\bullet\,]$
and $\hat{H}$, we can represent the action of $\mathcal{S}$ on $\mathcal{U}$
by a corresponding action on $\mbox{su}(\dim\mathcal{H})$. By
linearity, this must be a linear and invertible map $S$. Indeed,
from (\ref{eq:liemorph}) it follows that $S([\hat{H},\hat{H}'])=[S(\hat{H}),S(\hat{H}')]$:
$S$ is a Lie automorphism. A theorem in \cite{ref:LieAlgebras-Jacobson}
states that all automorphisms on $gl(n)$ (the general matrix algebra) are of the form $S(X)=AXA^{-1}$
or $S(X)=-AX^{T}A^{-1}$. Our automorphism is instead on the sub-algebra
$\mbox{su}(\dim\mathcal{H})$. By choosing a hermitian basis of $gl(n)$
we can extend it uniquely to one of $gl(n)$ and apply the theorem.
The additional hermitian structure demands furthermore that $A^{-1}=A^{\dagger}$.
Thus $S(\hat{H})=U\hat{H}U^{\dagger}$ or $S(\hat{H})=-U\hat{H}^{T}U^{\dagger}$.
The latter is excluded because it would not preserve the trace of
quantum states. Hence, under the premise of controllability, two systems
are indistinguishable if and only if they are related through a unitary
transformation \begin{gather*}
H_{k}=U\hat{H}_{k}U^{\dagger},\quad M_{\ell}=U\hat{M}_{\ell}U^{\dagger},\quad\rho_{0}=U\hat{\rho}_{0}U^{\dagger}.\end{gather*}

\paragraph*{Usage of a priori knowledge:---}

In practice, it is reasonable to assume that some elements of the
black box are known. Each known element shrinks the set of possible
unitary transformations, because, for example, $H_{k}=U\hat{H}_{k}U^{\dagger}=\hat{H}_{k}$
implies $[U,\hat{H}_{k}]=0$.

As an example, we consider two qubits coupled by an unknown Hamiltonian.
We estimate them by performing arbitrary operations $\hat{H}_{1}=X_{1}\otimes\openone_{2}$
and $\hat{H}_{2}=Y_{1}\otimes\openone_{2}$ on the first qubit and
by measuring $\text{a)}\ Z_{1}\otimes\openone_{2},$ $\text{b)}\ Z_{1}\otimes Z_{2},$
or $\text{c)}\ |\Psi^{-}\rangle_{12}\langle\Psi^{-}|,$ where $X_{i}$,
$Y_{i}$, and $Z_{i}$ are the Pauli operators of qubit $i=1,2$,
and $|\Psi^{-}\rangle_{12}=(|01\rangle_{12}-|10\rangle_{12})/\sqrt{2}$
is the singlet state. Assuming that the system is controllable, we
can apply the above results.

First, the conditions $[U,\hat{H}_{k}]=0$ reduce the unitary transformation
$U$ to $\openone_{1}\otimes U_{2}$, where $U_{2}$ is a unitary
operator acting on the second qubit, which may be parameterized as $U_{2}=e^{-\frac{i}{2}\theta\bm{n}\cdot\bm{\sigma}_{2}}$
with a unit vector $\bm{n}$. We then impose another condition $[U,\hat{M}]=0$:

a) In the first case with $\hat{M}=Z_{1}\otimes\openone_{2}$, this
condition is already satisfied and the unitary transformation $U$
is not reduced any further, $U=\openone_{1}\otimes e^{-\frac{i}{2}\theta\bm{n}\cdot\bm{\sigma}_{2}}$,
where remain three parameters.

b) In the second case with $\hat{M}=Z_{1}\otimes Z_{2}$, the condition
reduces $U$ to $\openone_{1}\otimes e^{-\frac{i}{2}\theta Z_{2}}$
with a single parameter.

c) Finally, in the third case with $\hat{M}=|\Psi^{-}\rangle\langle\Psi^{-}|$,
we have
%\begin{align*}
% & [U,\hat{M}]=-i\sin\frac{\theta}{2}\,\Bigl[-n_{z}(|\Psi^{+}\rangle\langle\Psi^{-}|-|\Psi^{-}\rangle\langle\Psi^{+}|)\\
% & +n_{x}(|\Phi^{-}\rangle\langle\Psi^{-}|-|\Psi^{-}\rangle\langle\Phi^{-}|)-in_{y}(|\Phi^{+}\rangle\langle\Psi^{-}|+|\Psi^{-}\rangle\langle\Phi^{+}|)\Bigr]
%\end{align*}
\begin{align*}
[U,\hat{M}]=i\sin\frac{\theta}{2}\,\Bigl(
n_{z}|\Psi^{+}\rangle\langle\Psi^{-}|
&-n_{x}|\Phi^{-}\rangle\langle\Psi^{-}|\\
&+in_{y}|\Phi^{+}\rangle\langle\Psi^{-}|
 \Bigr)+\text{h.c.},
\end{align*}
which is vanishing only when $\sin(\theta/2)=0$, i.e., $U=\openone$ up to an irrelevant phase.
This shows that the Bell measurement is more efficient to estimate
the system.

\paragraph*{Infection criterion for arbitrary systems:---}

Let us consider another more general example, a generic Hamiltonian
of a $d$-dimensional Hilbert space in the form \begin{equation}
H_{0}=\sum_{(n,m)\in E}c_{nm}|n\rangle\langle m|,\label{eq:hamm}\end{equation}
 where the orthonormal basis $|n\rangle$ may be thought of as ``local,''
and $E$ are the edges of the graph $G=(|n\rangle,E)$, that describes
the non-zero off-diagonal ($n\neq m$) couplings $c_{nm}$. We assume
that a set of nodes $C$ can be controlled ($H_{k}=|k\rangle\langle k|$,
$k\in C$), and that (at least) one particular node of $C$ can be
measured $(M_{1}=|1\rangle\langle1|$, $1\in C$). The crucial assumption
about the set $C$ is that it is ``infecting'' $G$ \cite{Burgarth2009a}.
This property is defined by the following propagation rules: 1) $C$
is ``infected''; 2) infected nodes remain infected; and 3) the
infection propagates from an infected node to a ``healthy'' neighbor
iff it is its only healthy neighbor. For an arbitrary Hamiltonian
we can always find an infecting set; how many nodes it contains depends
on how sparse the Hamiltonian is in the particular basis of consideration.
In practice there are physical choices of the basis corresponding
to local operations, and many Hamiltonians are infected by acting
on a vanishing fraction of nodes only.

Based on the assumption that $C$ is infecting one finds that the
system is controllable, so our theorem can be applied. Firstly, there
is a $k\in C$ that has a unique neighbor $\ell$ outside $C$. For
that $k$ we have $[iH_{k},iH_{0}]=-\sum_{m\in n(k)}(c_{km}|k\rangle\langle m|-c_{mk}|m\rangle\langle k|)$,
where $n(k)$ is the neighborhood of $k$. Commuting it with $iH_{k}$ again yields
\begin{equation}
\bm{[}[iH_{k},iH_{0}],iH_{k}\bm{]}=i\sum_{m\in n(k)}(c_{km}|k\rangle\langle m|+c_{mk}|m\rangle\langle k|).\label{eq:sum}\end{equation}
For $m\in n(k)\cap C$, on the other hand, we can single out terms
by \[
\bm{[}[iH_{m},iH_{0}],iH_{k}\bm{]}=-i(c_{km}|k\rangle\langle m|+c_{mk}|m\rangle\langle k|).\]
By adding these to (\ref{eq:sum}) for all $m\in n(k)\cap C$, only a single term $i(c_{k\ell}|k\rangle\langle\ell|+c_{\ell k}|\ell\rangle\langle k|)$ is left.
Commuting this with $iH_{k}$ again gives $c_{k\ell}|k\rangle\langle\ell|-c_{\ell k}|\ell\rangle\langle k|$.
Finally, commuting the latter two and subtracting the term proportional
to $iH_{k}$ we are left with $i|\ell\rangle\langle\ell|$. By induction,
we can obtain $|n\rangle\langle n|$, $\forall n$. This implies full
controllability \cite{Schirmer2001}.

If we assume that $H_{k}$ and $M_{1}$ are known, we need to
look at the unitaries that commute with these operators. There will
be many. However, we will assume here the \emph{knowledge} that the
Hamiltonian $H_{0}$ has the form given in (\ref{eq:hamm}). Hence,
we are talking about an indirect coupling strength estimation \cite{Burgarth2009a,Burgarth2009,Franco2009,Burgarth2011},
where the topology $E$ is known while the parameters are unknown.
Let us see what this knowledge implies. Firstly, we have to have $[H_{k},U]=0=-[H_{k},U^{\dagger}]$
$(k\in C)$. Since $H_{k}$ are projectors that implies that $|k\rangle$
must be an eigenstate of $U$ and $U^{\dagger}$ for all $k\in C$.
The estimated Hamiltonian $\hat{H}_{0}=UH_{0}U^{\dagger}$ has to
be of the form $\hat{H}_{0}=\sum_{(n,m)\in E}\hat{c}_{nm}|n\rangle\langle m|$,
where $\hat{c}_{nm}$ are unequal to zero and could in principle differ
from $c_{nm}$. The edges $E$ must be the same for both $H_{0}$
and $\hat{H}_{0}$ because we assume knowledge of the topology. Because
$C$ is an infecting set, there is one $k\in C$ that has a unique
neighbor $\ell$ outside of $C$. The corresponding term in the Hamiltonian
$H_{0}$ is $c_{kl}|k\rangle\langle\ell|+c_{lk}|\ell\rangle\langle k|$.
Because $|k\rangle$ is an eigenstate of $U$ this transforms under
$U\bullet U^{\dagger}$ into $c_{kl}e^{i\phi_{k}}|k\rangle\langle\ell|U^{\dagger}+c_{lk}e^{-i\phi_{k}}U|\ell\rangle\langle k|$.
Because the edges $E$ are the same for $H_{0}$ and $\hat{H}_{0}$
there is a corresponding term $\hat{c}_{kl}|k\rangle\langle\ell|+\hat{c}_{lk}|\ell\rangle\langle k|$
in $\hat{H}_{0}$. Furthermore, since $|k\rangle$ is an eigenstate
of $U^{\dagger}$ no other node $|n\rangle$ can be brought to $|k\rangle$,
i.e., $\langle k|U|n\rangle=0$. Given that $\ell$ is the only node
outside $C$ coupled to $k$ we conclude \[
c_{kl}e^{i\phi_{k}}|k\rangle\langle\ell|U^{\dagger}+c_{lk}e^{-i\phi_{k}}U|\ell\rangle\langle k|=\hat{c}_{kl}|k\rangle\langle\ell|+\hat{c}_{lk}|\ell\rangle\langle k|,\]
 which implies that $|\ell\rangle$ is an eigenstate of $U$. Finally,
by induction we get that $U$ must be a diagonal matrix in the {}``local''
basis $|n\rangle$. Thus, up to the local phases of the basis vectors
the Hamiltonian $H_{0}$ is uniquely estimated. What is remarkable
here is that we do not have to assume the knowledge of the phases
of $c_{nm}$ and it suffices to measure a single node. This generalizes
the previous results \cite{Burgarth2009a,Burgarth2009,Franco2009,Burgarth2011}
substantially.

\paragraph*{Conclusion:---}

We have shown that controlable closed quantum systems can be estimated,
in principle, \emph{up to unitary conjugation. }This provides an easy-to-check
criterion for experimental setups, telling us which extra controls,
measurements or a priori knowledge are needed to achieve highly accurate
quantum system identification required for quantum computation. We
have applied this criterion to a simple setup to show how the structure
of the measurement observables can be important for estimation efficieny,
and we have constructed a method to obtain a fully controllable \emph{and
estimable} system from an arbitray Hamiltonian.

\begin{acknowledgments}
We acknowledge fruitful discussions with Madalin Guta, Koenraad Audenaert,
Vittorio Giovannetti, Koji Maruyama, and Martin B. Plenio. This work
is supported by a Special Coordination Fund for Promoting Science
and Technology, and a Grant-in-Aid for Young Scientists (B),
both from MEXT, Japan, and by the EPSRC grant EP/F043678/1.
\end{acknowledgments}

\end{document}